\documentclass[]{mn2e}
\usepackage{amssymb}
\usepackage{amsmath}
\newif\ifAMStwofonts
\def\B{\bmath{B}}
\def\J{\bmath{J}}
\def\E{\bmath{E}} 
\def\vr{v_{r}}
\def\vf{v_{\phi}}            
\def\vk{v_{K}}               
\def\vz{v_{z}}               
\def\ee #1 {\times 10^{#1}}
\input epsf

\title{Angular momentum transport in protostellar discs}
\author[R. Salmeron, A. K\"onigl and M. Wardle]
       {Raquel Salmeron$ ^1 $, Arieh K\"onigl$ ^1 $ 
\& Mark Wardle$ ^2 $ \\
$ ^1 $Department of Astronomy \& Astrophysics, The University of
Chicago, Chicago  IL 60637, USA \\
$ ^2 $Physics Department, Macquarie University, Sydney NSW 2109, Australia}

\date{2006 November 6} 
\pagerange{\pageref{firstpage}--\pageref{lastpage}}
\pubyear{2006} 
\begin{document} 
\maketitle 
\label{firstpage}
\begin{abstract} 
Angular momentum transport in protostellar discs can take place either
radially, through turbulence induced by the magnetorotational
instability (MRI), or vertically, through the torque exerted by a
large-scale magnetic field that threads the disc. Using semi-analytic
and numerical results, we construct a model of steady-state discs that
includes vertical transport by a centrifugally driven wind as well
as MRI-induced turbulence. We present approximate criteria for the
occurrence of either one of these mechanisms in an ambipolar
diffusion-dominated disc. We derive ``strong field'' solutions in which
the angular momentum transport is purely vertical and ``weak field''
solutions that are the stratified-disc analogues of the
previously studied MRI channel modes; the latter are transformed
into accretion solutions with
predominantly radial angular-momentum transport when we implement a
turbulent-stress prescription based on published results of numerical
simulations.  We also analyze ``intermediate field strength'' solutions
in which both modes of transport operate at the same radial
location; we conclude, however, that significant spatial overlap of these two
mechanisms is unlikely to occur in practice.
To further advance this study, we have developed a general scheme that
incorporates also the Hall and Ohm conductivity regimes in discs with a
realistic ionization structure.
\end{abstract}

\section{Introduction}
\label{sec:intro}
 
The mechanisms responsible for transporting angular momentum in
protostellar discs are still not well understood. Perhaps the most
generally relevant processes are:

($a$) Radial transport through turbulence induced by the
magnetorotational instability (MRI; e.g. Balbus \& Hawley 1998). The MRI
acts, essentially, by converting the free energy of differential
rotation into turbulent motions that transfer angular momentum radially
outward via the Maxwell stress of small-scale, disordered magnetic
fields.

($b$) Vertical transport by outflows driven centrifugally from the disc
surfaces (e.g. K\"onigl \& Pudritz 2000). The transfer of angular
momentum from the inflowing gas to the wind is mediated by a
large-scale, ordered magnetic field that threads the disc (likely
corresponding to interstellar field lines advected inward by the
accretion flow but possibly generated locally by a disc dynamo).

Radial angular-momentum transport has been the hallmark of generic
viscous disc models, for which the MRI process is thought to provide the
most cogent physical basis. On the other hand, the ubiquity of bipolar
outflows and jets in protostars and the apparent correlation between
outflow and accretion signatures in these systems (e.g. Hartigan et
al. 1995) have led to the suggestion that disc outflows might be a key
ingredient of the accretion process on account of the high
angular-momentum transport efficiency of centrifugally driven winds
(e.g. K\"onigl 1989). The possibility of a central role for vertical
angular-momentum transport is supported by recent {\it HST}\/
observations of protostellar jets, which have been interpreted in terms
of magnetocentrifugal winds carrying a large fraction of the excess
angular momentum of the accreted matter from the inner regions ($r
\lesssim 1\;$AU) of the associated discs (see Ray et al. 2006 for a
review). Although thermal pressure may contribute to the driving of the
outflows (e.g. Pesenti et al. 2004), the stresses exerted by a
large-scale magnetic field dominate the flow acceleration and
collimation (Blandford \& Payne 1982, hereafter BP82; see, however,
Soker \& Regev 2003 for an alternative view of the importance of thermal
acceleration) and are the key to the efficient vertical angular-momentum
transport indicated by the {\it HST}\/ measurements.

Previous analytic and numerical investigations have considered the above
two mechanisms under various simplifying assumptions. MRI studies have
examined both the linear and nonlinear stages of the instability
(e.g. Sano \& Stone 2002a,b; Salmeron \& Wardle 2003, 2005). The
long-term global evolution of the instability in vertically stratified
discs that are threaded by a single-polarity magnetic field is still an
open question (e.g. Miller \& Stone 2000). The study of wind-driving
discs has been complicated by the need to solve simultaneously for the
inflow and outflow structures and by the non-local nature of the MHD
solution. Progress has nevertheless been made semi-analytically by
adopting a self-similarity formulation (e.g.  Wardle \& K\"onigl 1993,
hereafter WK93; Li 1995, 1996; Ferreira 1997) as
well as numerically, with recent simulations reporting an evolution to a
quasi-stationary state (e.g. Casse \& Keppens 2002; Kuwabara et
al. 2005). Since protostellar discs are weakly ionized over most of
their extent, magnetic diffusivity effects have to be taken into account
in modeling the accretion flow in both cases. In particular, both the
MRI-driven turbulence and the centrifugal wind-launching mechanism
require a minimum level of field--matter coupling to be effective
(e.g. WK93; Sano \& Inutsuka 2001). Therefore, the detailed ionization
structure and conductivity properties of the disc need to be determined
in order to incorporate either one of these processes into a realistic
model.

Angular momentum removal from real discs is likely to involve both the
radial and vertical transport mechanisms discussed
above.\footnote{Vertical angular momentum transport could alternatively
take place through the process of magnetic braking, which involves the
propagation of torsional Alfv\'en waves into the surrounding
interstellar medium. This possibility is neglected in what follows, but
see Krasnopolsky \& K\"onigl (2002).} There have been a few attempts in
the literature to construct quasi-steady disc models in which both of
these mechanisms operate (e.g. Lovelace et al. 1994; Casse \& Ferreira
2000; Ogilvie \& Livio 2001), but these treatments invariably used a
prescription for radial angular-momentum transport that did not
explicitly account for its origin in MRI-induced turbulence. However,
the fact that disc regions in which the entire excess angular momentum
is carried away by a centrifugally driven wind are generally stable to
the fastest growing mode of the MRI (WK93) indicates that a more refined
approach is required. Broadly speaking, the MRI is suppressed in regions
where the ratio $a\equiv v_{{\rm A}z}/c_{\rm s}$ of the Alfv\'en speed
corresponding to the vertical field component $B_z$ and the isothermal
sound speed is comparatively large ($\gtrsim 1$); it may be expected to
evolve into a fully developed turbulence when $a \ll 1$ (provided also
that the gas is well coupled to the field). Since the value of $a$
increases with distance $z$ from the disc midplane, these considerations
suggest that it could in principle be possible for both vertical and
radial angular-momentum transport to take place at the same disc radius
$r$, with the radial transport confined to a given range in $z$ and with
the vertical transport dominating at greater heights.

The purpose of this paper is to construct semi-analytic models of
protostellar discs that incorporate both centrifugally driven winds and
MRI-induced turbulence and to identify the relevant parameter
regimes where either one of these two mechanisms dominates the
angular momentum transport. 
We pay particular attention to the possibility that both of
these mechanisms operate at the same disc radius and consider how they
might affect each other in that case. We describe our wind-driving disc
model (formulated following WK93) and present an illustrative solution
of a disc with a predominantly vertical angular momentum transport in
Section 2. In Section 3 we outline our method of including MRI-induced
turbulence (taking account of recent numerical simulations of the
nonlinear evolution of the MRI) and illustrate it with a solution of a
disc with a predominantly radial angular-momentum transport. We then
consider the case of combined transport by wind and turbulence in
Section 4 and summarize our results in Section 5.

\section{Vertical Transport}
\label{sec:vertical}

We have devised a general scheme for modeling wind-driving protostellar
discs using a tensor-conductivity formalism and a realistic vertical
ionization structure.\footnote{The main ionization sources outside the 
innermost $\sim 0.1\;$AU from the central object are nonthermal: interstellar 
cosmic rays, X-ray and UV radiation emitted by the magnetically active
protostar and the decay of radioactive elements present in the
disc. The gas near the disc surface is ionized primarily by X-rays in
the vicinity of the star, and by UV photons (with a contribution also
from the interstellar radiation field) in the outer disc (e.g. Semenov 
et al. 2004; Glassgold et al. 2005).} The conductivity 
tensor makes it possible to
include in a systematic way the three relevant magnetic diffusion
mechanisms (ambipolar, Hall and Ohmic). The solutions presented in this
paper are, however, derived under the simplifying assumptions of pure
ambipolar diffusivity and (except in Section 4) constant ion
density. This enables us to apply the analytic results obtained for this
case in WK93 to the formulation of a prescription for MRI-induced
turbulent transport in a wind-driving disc model. A wider class of
solutions of our general scheme will be presented in future
publications.

Our wind solutions are based on the magnetocentrifugal mechanism for
``cold'' disc outflows proposed by BP82. In this scenario, matter at the
disc surface is accelerated by magnetic stresses if the open magnetic
field lines that thread the disc are inclined at a sufficiently large
angle to the rotation axis (the ``bead-on-a-wire'' effect), potentially
reaching super-Alfv\'enic speeds. Our modeling procedure is based on
that of WK93 and we refer the reader to that study for further
details. In brief, we solve the mass and momentum conservation equations 
for the neutral gas, and the induction equation for
the evolution of the magnetic field, assuming a steady-state,
geometrically thin and nearly Keplerian accretion disc and neglecting
all radial derivatives except that of $v_\phi$.  The disc is further
assumed to be isothermal and weakly ionized. The latter condition
implies that ionization and recombination processes, as well as the
inertia and thermal pressure of the charged particles, have a negligible
dynamical effect.  In this limit, separate equations for the ionized
species are not needed, and their effect on the neutrals can be conveyed
by means of a conductivity tensor $\bmath{\sigma}(r,z)$ (assuming
cylindrical polar coordinates; see Wardle 1999 and references
therein). In addition, the field must satisfy $\nabla \cdot \B = 0$
(which in the thin-disc approximation implies that $B_z$ can be assumed
not to vary with height in the disc), whereas the current density $\J$
must satisfy Amp\`ere's and Ohm's laws.

The foregoing equations are integrated vertically from the midplane
(indicated by a subscript $0$) up to the position $z_{\rm s}$ of the
critical (sonic) point, and the values of the physical variables there
are estimated. The solution is then integrated backward to a fitting
point and iterated until full convergence is reached. The resulting disc
solution is matched on to a global BP82-type self-similar wind solution,
obtained by imposing the regularity condition at the Alfv\'en critical
point. This final step effectively determines the value of $B_\phi$ at
the disc surface (subscript b, defined as the height where $v_{\phi}$
equals the Keplerian speed $v_{\rm K}$).  The overall solution is
determined by the following parameters:
\begin{enumerate} 
\item $a_0 = v_{{\rm A}z,0}/c_{\rm s}$, the midplane ratio of the Alfv\'en
speed (based on the uniform vertical field component) to the sound
speed, which measures the magnetic field strength.
\item $\eta$, the ratio of the Keplerian rotation time $\Omega_{\rm
K}^{-1}=r/v_{\rm K}$ to the neutral--ion momentum-exchange time (see
WK93); this parameter measures the degree of coupling between the
neutrals and the ionized gas component (and magnetic field) and does not
vary with height if the ion density is
constant.\footnote{\label{coupling}In the more general formulation, one
needs to specify the values of the independent components of the
conductivity tensor -- viz. the field-aligned ($\sigma_{\parallel}$),
Pedersen ($\sigma_{\rm P}$) and Hall ($\sigma_{\rm H}$) conductivities.
In this case, the magnetic coupling is expected to \emph{decrease} with
height in the ambipolar diffusion-dominated upper section of the disc,
where the ion density is most strongly affected by the decrease of the
neutral gas density with $z$ (e.g. Salmeron \& Wardle 2005). In the
simplified approach adopted in this paper we use the parameter $\eta =
(4\pi c_{\rm s}^2/c^2) a^2 (\sigma_{\rm P}/\Omega_{\rm K})$ instead
(with $a \propto \rho^{-1/2}$ increasing with $z$) and (except in
Section 4) assume that it remains constant with height.}
\item $\epsilon \equiv -v_{r0}/c_{\rm s}$, the normalized inward radial
speed at the midplane. In practice, the value of this parameter is
determined in our solutions (for given values of the other parameters)
by imposing the Alfv\'en critical-point constraint on the wind.
\item $c_{\rm s}/v_{\rm K} = h_{\rm T}/r$, the ratio of the tidal scale
height to the radius -- a measure of the geometric thinness of the disc.
\item $\epsilon_B \equiv -cE_{\phi}/c_{\rm s} B_z$, the normalized azimuthal
component of the electric field $\E$. This parameter vanishes in a strictly
steady-state solution but is nonzero if the magnetic field lines drift
radially. Our formulation remains valid if the field--line drift speed is 
small compared to $|v_r|$ (see WK93).
\footnote{\label{ansatz}Our
ansatz $\epsilon_{\rm B} = 0$ effectively fixes the value of $B_r$ at the disc
surface. In a generalization to a global treatment, $B_{r,{\rm b}}$ would be
determined by the radial distribution of $B_z$ along the disc (e.g.
Ogilvie \& Livio 2001; Krasnopolsky \& K\"onigl 2002), and the value of
$\epsilon_{\rm B}$ could be inferred from the overall magnetic field
structure and the disc conductivity.}
\end{enumerate}
\begin{figure}
\centerline{\epsfxsize=7.5cm \epsfbox{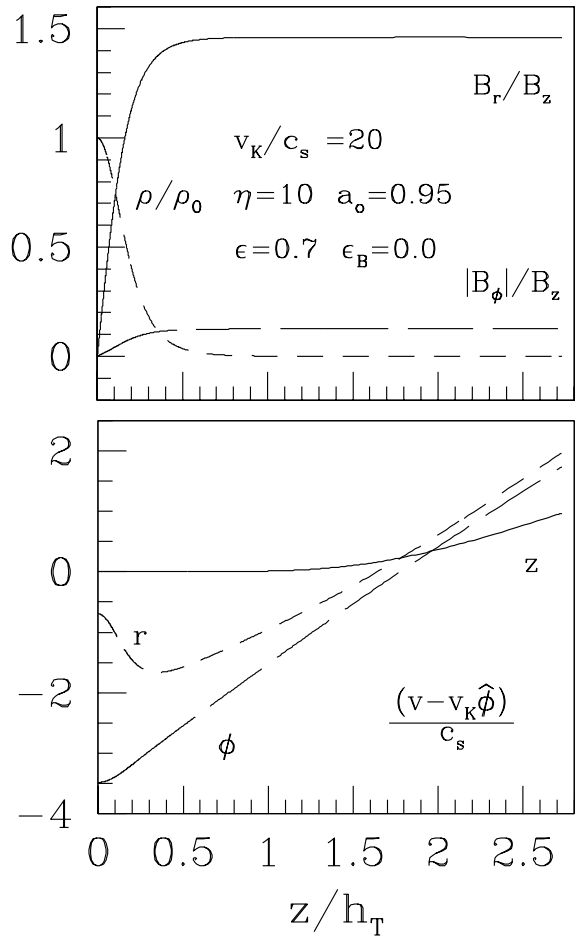}}\vskip 0cm
\caption{Vertical structure of a strongly magnetized, wind-driving
disc. {\it Top:}\/ $B_r$, $|B_\phi|$, and $\rho$. {\it
Bottom:}\/ velocity components. The curves terminate at the
sonic point. The disc solution parameters are indicated in the
figure. The corresponding parameters of the BP82 self-similar wind solution are
$\kappa = 3.2 \ee -4 $ (normalized mass-to-magnetic flux ratio),
$\lambda = 395$ (normalized total specific angular momentum), and
$\xi_{\rm b}^{\prime} \equiv B_{r,{\rm b}}/B_z = 1.46$.}
\label{fig:wind} 
\end{figure}

Fig.~\ref{fig:wind} shows an example of a local disc solution that matches
on to a global BP82 wind solution. 
This solution resembles the one presented in Fig.~10 of WK93, although
it was derived for a different choice of the parameters $\eta$ and
$a_0$. Both of these solutions illustrate the main properties of
wind-driving, strongly magnetized ($a_0 \lesssim 1$) discs. In the
quasi-hydrostatic layer just above the midplane the magnetic field lines
are (radially) bent and (azimuthally) sheared and the field removes
angular momentum from the matter. The scale height $h$ in this region is
reduced by magnetic squeezing, which dominates the tidal
compression. Above this layer lies a transition region where the
magnetic energy comes to exceed the gas internal energy and the field
lines are locally straight. In both of these zones, the fluid azimuthal
speed is sub-Keplerian. Since the gas angular velocity decreases with
radius, the field lines (which move with a constant angular velocity)
eventually overtake the matter and fling it out centrifugally. The
region where $v_r$ and ($v_\phi-v_{\rm K}$) become positive constitutes
the base of the wind. It has been shown (WK93; K\"onigl 1997) that solutions 
of this type satisfy 
\begin{equation}
(2\eta)^{-1/2} \lesssim a_0 \lesssim \sqrt{3}  \lesssim \epsilon\eta 
\lesssim v_{\rm K}/2 c_{\rm s}\ .
\label{eq:ineq}
\end{equation}
The first inequality in equation~(\ref{eq:ineq}) expresses the
requirement that the fluid remain sub-Keplerian below the
wind-launching region, the second gives the wind launching condition
($B_{r,b}/B_z > 1/\sqrt{3}$), and the third gives the
constraint that the base of the wind be located above a density scale
height. The first two conditions imply a minimum value for the
field--matter coupling parameter $\eta$ (which a more detailed analysis
sharpens to $\eta > 1$), whereas the last two imply that magnetic
squeezing dominates the vertical confinement of the disc ($h<h_T$). The
last inequality specifies that the rate of ambipolar diffusion heating
at the midplane should not exceed the rate of release of gravitational
potential energy.

\section{Radial Transport}
\label{sec:radial}

Numerical simulations indicate that vigorous radial angular momentum
transport is sustained by the MRI when the initial (subscript i)
Elsasser number $\Lambda_{\rm i} \equiv v_{{\rm A}z,{\rm
i}}^{2}/\eta_{\rm Ohm}\Omega_{\rm K}$ is $\gtrsim 1$ (e.g. Sano \&
Inutsuka 2001; Sano \& Stone 2002a,b), where $v_{{\rm A}z,{\rm i}}$ is
the (midplane) Alfv\'en speed based on the initial vertical field
$B_{z,{\rm i}}$ and $\eta_{\rm Ohm}$ is the Ohmic
diffusivity.\footnote{$\Lambda$ is often referred to in the MRI
literature as Re$_{\rm M}$, the magnetic Reynolds number, but in general
${\rm Re_M} \ne \Lambda$.} Now, for the vertical initial field geometry
assumed here, the currents and fields are mutually orthogonal and
ambipolar diffusion acts as a field-dependent Ohmic resistivity
(e.g. Balbus \& Terquem 2001). Under these conditions one can show that
$\Lambda_{\rm i} = \eta_{\rm i}$, so the minimum-coupling condition in
this case is the same for vertical and for radial angular-momentum
transport.

\begin{figure}
\centerline{\epsfxsize=7.5cm \epsfbox{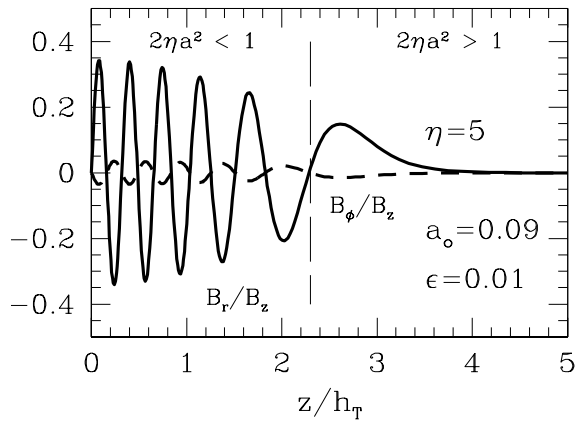}}\vskip 0cm
\caption{Structure of a weakly magnetized disc with vertical magnetic
angular-momentum transport in the limit where no
wind develops (channel solution). The dashed line marks the nominal
boundary of the MRI-unstable zone discussed in the text.}
\label{fig:wiggles}
\end{figure}

What determines which mode of angular momentum transport dominates when
the coupling is sufficiently strong ($\eta \gtrsim 1$)?
As already noted in Section 1, the MRI is suppressed for sufficiently
high values of the field-strength parameter $a$. A useful clue to the
lower limit for MRI stability is provided by the first inequality in
equation~(\ref{eq:ineq}). As discussed in WK93, when this inequality is
violated the surface layers of the disc become super-Keplerian, implying
outward streaming motion that is unphysical in the context of a pure
wind-driving disc model. However, as elaborated on below, an outward motion
of part of the disc material is a characteristic of the two-channel
MRI mode that underlies MRI-induced turbulence. Further support for this
choice of an upper limit on $a$ in the MRI-unstable region is provided by
the fact that wind-driving disc solutions that satisfy the constraints
expressed by equation~(\ref{eq:ineq}) are {\em stable} to
the fastest growing linear mode of the MRI, essentially because the
minimum wavelength of this mode exceeds the magnetically reduced disc
scale height (see WK93).\footnote{Note, however, that the formal
demonstration of this fact requires using only the third inequality in
equation~(\ref{eq:ineq}).} The above limit on $a$ was applied by WK93 at
the midplane, since they considered discs in which the entire excess
angular momentum of the  accretion flow is transported by a wind. As we
are now concerned with the possible existence of an alternative mode of
transport, we propose to generalize this condition and apply it {\em
locally} (i.e. using $a(z) \propto \rho(z)^{-1/2}$ instead of its midplane
value) to differentiate the MRI-unstable section of the disc (where
$2\eta a^2 < 1$) from the region where only vertical angular
momentum transport takes place ($2\eta a^2 > 1$).

By integrating the disc equations of the model described in Section 2
for the case in which $a<(2\eta)^{-1/2}$ over the bulk of the disc we
have verified that such systems do not develop winds.  Since our current
formulation does not allow for an alternative means of transporting
angular momentum through the disc surfaces (such as magnetic braking,
``failed'' winds that do not cross the relevant critical surfaces, or
nonsteady phenomena), the only remaining option is for the angular
momentum to be transported magnetically between different heights and
then to be carried away radially by bulk flow (with the gas that loses
angular momentum moving in and the material that gains angular momentum
moving out). Fig.~\ref{fig:wiggles} shows a solution of this type. The
radial field component oscillates in the region where $2\eta a^2
\lesssim 1$, with $B_\phi$ exhibiting corresponding oscillations with an
opposite sign. The regions where $B_z \partial B_\phi/\partial z < 0$
lose angular momentum to the field and have $v_r <0$, with the converse
behaviour characterizing regions where $B_z \partial B_\phi/\partial
z>0$. We associate this solution with a two-channel MRI mode for the
case of a stratified, non-ideal MHD disc.\footnote{Similar field
configurations were obtained by Ogilvie \& Livio (2001), who also
related them to the channel solution.} As was first shown by Goodman \&
Xu (1994), such modes (which they considered in the unstratified,
ideal-MHD case) are exact solutions in both the linear and the nonlinear
regimes.  Their analysis, however, revealed that these solutions are
{\em unstable} to parasitic modes, and subsequent numerical simulations
(carried out in both the ideal and non-ideal regimes) have verified that
they rapidly evolve into a turbulent state (e.g. Hawley et al. 1995;
Sano et al. 2004, hereafter S04). 

To account for the turbulent angular momentum transport that we expect
to develop in disc regions where $\eta\gtrsim 1$ and $2a^2\eta\lesssim 1$,
we adopt the results of numerical simulations, which have found
that the main contribution to the $r\phi$ component of the stress tensor is 
provided by the turbulent Maxwell stress and that its spatial and
temporal average $\ll w_{r\phi} \gg$ can be expressed in terms of the
similarly averaged magnetic pressure by
\begin{equation}
\ll w_{r\phi} \gg\,  \approx 0.5 \, \frac{\ll B^2 \gg}{8 \pi}
\label{eq:wrphi}
\end{equation}  
(e.g. Hawley et al. 1995; S04). Using the characteristic ratios
$\ll B_{\phi}^2 \gg/\ll B_z^2 \gg\,  \approx 24$ and $\ll B_r^2 \gg/\ll
B_z^2 \gg\,  \approx 3$ for MRI turbulence reported in Table 4 of S04, we
substitute $\ll B^2 \gg \, \approx 28 \ll B_z^2 \gg$ into equation
(\ref{eq:wrphi}) to get
\begin{equation}
\label{eq:mri}
\ll w_{r\phi} \gg \approx 14\,  Y\,  \frac{B_{z,{\rm i}}^2}{8 \pi}\ ,
\end{equation}
where $Y \equiv \, \ll B_z^2 \gg /B_{z,{\rm i}}^2 =
Y(\beta)$ is listed in Table 2 of S04 for a range of isothermal, 
uniform-$B_{z,{\rm i}}$ models as a function of $\beta_{\rm i} \equiv
2/a_{\rm i}^2$, the 
initial plasma $\beta$ parameter. As the values of $\beta$ in our
applications are smaller (by up to $\sim 1-2$ orders of magnitude at the
midplane) than the ones listed in S04, we resort to extrapolation, which
can be justified by the relatively weak dependence of $Y$ on $\beta$:
we obtain $Y= 2$, $7$ and $10$ for $\beta = 10$, $100$ and $500$,
respectively.
\begin{figure}
\centerline{\epsfxsize=7.5cm \epsfbox{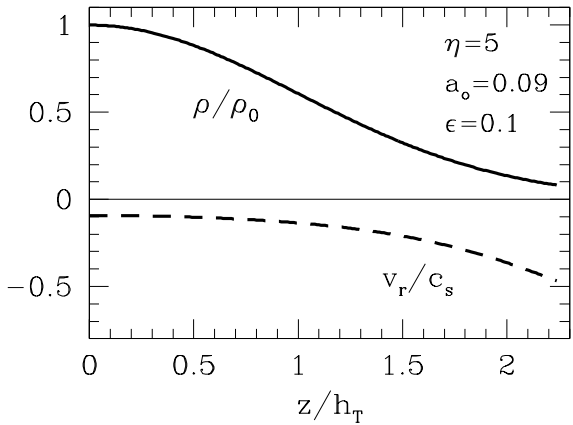}}\vskip 0cm
\caption{Structure of a weakly magnetized disc with purely radial
angular-momentum transport given by the MRI-induced turbulence
prescription discussed in the text. The curves terminate at the point
where the parameter $a$ satisfies $2\eta a^2=1$.}
\label{fig:radial}
\end{figure}

Radial angular-momentum transport is incorporated into the disc
angular momentum conservation equation by adding (in the relevant disc
region) the
turbulent stress term, which we do by approximating $(1/r)\partial (r^2 \ll
w_{r\phi} \gg )/ \partial r \approx \, \ll  w_{r\phi}\gg$. This equation
then reads
\begin{equation} 
\label{eq:momentum} 
\frac{\rho \vr \vk}{2r} + \rho \vz
\frac{d \vf}{dz} = -\frac{J_r B_z}{c} - \frac{\ll w_{r\phi}\gg}{r}\ ,
\end{equation} 
where we used $\partial(r \vf)/\partial r \approx \vk/2$ and $J_z
\approx 0$. We assume that the small-scale (turbulent) magnetic field
components only contribute to the radial angular-momentum transport. We
interpret all the other magnetic-field (and velocity) terms that appear
in the disc structure equations as corresponding to the {\em mean} values
of the respective quantities. In a similar vein, we interpret the first
term on the r.h.s. of equation~(\ref{eq:momentum}) as representing the
vertical angular-momentum transport induced by the mean field.
This representation is akin to that of previous models that incorporated radial
angular-momentum transport into discs threaded by an open magnetic
field (see Section 1). The underlying assumptions of this approach --
that the relationships obtained from local shear-box simulations remain
applicable in a stratified disc, that $\ll B_r B_z\gg$ averages
out to zero and that $\ll (B_r^2 + B_\phi^2)\gg$ does not have a
significant $z$ dependence -- will, however, need to be verified by
means of global numerical simulations.
According to the above formulation, vertical angular-momentum transport
associated with a large-scale, ordered field occurs throughout the disc
(with the angular momentum deposited into the mean field
eventually transferred back to matter in the wind region above the
disc), whereas radial transport effected by MRI-induced turbulence is
confined to the layer where the conditions $1 \lesssim \eta \lesssim
(2a^2)^{-1}$ are satisfied.

Fig.~\ref{fig:radial} shows a weak-field disc solution, obtained by
implementing the above prescription for the same values of $\eta$ and
$a_0$ as in the channel solution presented in
Fig.~\ref{fig:wiggles}. In this case the angular momentum transport is 
purely radial throughout the disc. The azimuthal field component at the 
top of the disc does not generally vanish in weak field disc configurations
of these type, implying a net vertical
torque ($\propto B_z B_{\phi,{\rm b}}$) on the disc.
However, in the current formulation such 
a torque can only be included if the wind launching condition is
satisfied. Since this is not the case here, we chose to eliminate all
the magnetic terms from the derived solution for self
consistency. The resulting error is, however, small since the
mean-field contribution to the disc dynamics (in particular, to angular
momentum transport and to vertical squeezing) is rather minor in this
instance. The lack of strong magnetic squeezing is reflected in the more
gradual decline of the density with height in this solution as compared
with the pure wind-driving disc model presented in Fig.~\ref{fig:wind}.

\section{Combined Transport}
\label{sec:combined}

We now consider the case of a disc with moderate coupling ($\eta \sim 1$)
and field strength ($a_0 < (2\eta)^{-1/2}$ but 
$a \sim 1$ at the surface),  so both radial angular-momentum transport
by a turbulent magnetic field (between the midplane and the height $z_{\rm t}$ 
where $2\eta a^2 \approx 1$) and vertical transport by a large-scale field 
(between $z=0$ and the top of the disc at $z_{\rm b}$) take place at the
same radial location. Solutions of this type occupy a limited region of
parameter space, corresponding to a large enough $a_0$ for a wind to be
launched from the disc surfaces and a sufficiently small $\eta_0$ to
render $2 \eta_0 a_0^2 < 1$ (but $\eta_0 \gtrsim 1$ so as not to
suppress the growth of MRI-unstable modes). We have found that if $\eta$
is assumed to be both $>1$ and constant with height (as was done in the
previous Sections), the vertical extent ${\Delta z}_{\rm MRI}$ over
which radial transport operates is very narrow. The value of ${\Delta
z}_{\rm MRI}$ is limited by the fact that $a_0$ has a lower limit below
which the wind-driving surface layers of the disc become sub-Keplerian
(corresponding to an effective violation of the first inequality in
equation~\ref{eq:ineq} in these regions). To circumvent this limitation,
we have allowed $\eta$ to fall slightly below 1 (with the expectation
that it is still large enough to sustain MRI turbulence; see Sano \&
Stone 2002a,b) and to decrease with height. The adopted profile is
motivated by the inferred behaviour of $\eta(z)$ in the ambipolar
diffusion-dominated upper layers of realistic protostellar discs at
$r\sim 1-10\;{\rm AU}$, where the ion density is most strongly affected
by the decrease of the neutral gas density with height (see
footnote~\ref{coupling}).
\begin{figure}
\centerline{\epsfxsize=7.5cm \epsfbox{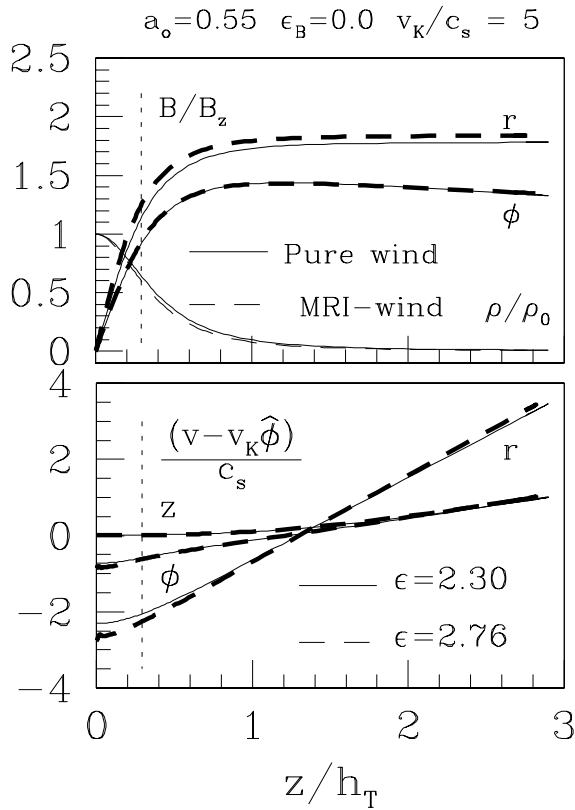}}\vskip 0cm
\caption{Wind-driving disc with a moderately strong field ($a_0 = 0.55$) and 
$\eta$ decreasing from $0.63$ to $0.5$ between the midplane and the 
surface. 
Density, magnetic field components ({\it top}) and velocity components 
({\it bottom}) are shown for
a solution that only incorporates vertical angular-momentum 
transport (solid lines) and also for a case where radial transport is 
included (dashed lines) in the region where $2\eta a^2 < 1$ (to the left of 
the vertical dotted line).
The curves terminate at the respective sonic points.}  
\label{fig:mri} 
\end{figure}

Fig.~\ref{fig:mri} shows illustrative solutions obtained under these
assumptions. The solid lines depict the structure of the disc when only
vertical angular-momentum transport is included
(cf. Fig.~\ref{fig:wind}).  The dashed lines indicate how the above
solution is modified when radial transport is incorporated using the
prescription discussed in Section 3 and keeping the value of $a_0$
fixed. Key properties of the two solutions are compared in
Table~\ref{table:boundary}, assuming that $c_{\rm s}$ and $\rho_0$ also
remain unchanged. The mass accretion rate per disc circumference is
evaluated from $\dot M_{\rm in}/2\pi r_0 = - 2 \int_0^{z_{\rm b}}{\rho
v_r dz}$; the listed value ($\dot{\cal{M}}_{\rm in}$) is this quantity
normalized by $(\rho_0 c_{\rm s} h_{\rm T})$.  The vertical and radial
torques (per unit area) on the disc are normalized by $r_0 B_0^2/4\pi$
and are given by $T_z = \kappa \lambda$ and $T_r = (4\pi{\ll
w_{r\phi}\gg}/B_0^2)({\Delta z}_{\rm MRI}/r_0)$, respectively (with
${\Delta z}_{\rm MRI} = z_{\rm t}$ for the given setup), whereas the
azimuthal field component at the disc surface is obtained from
$|B_{\phi,{\rm b}}|/B_z = \kappa (\lambda - 1)$. The heights $z_{\rm t}$,
$z_{\rm b}$, and $z_{\rm s}$ are listed in units of the scale height
$h$, which is determined for each solution as the value of $z$ where the
density drops to $\rho_0/\sqrt{e}$.

The addition of radial angular-momentum transport results in this case
in an increased inflow speed (measured by $\epsilon$) and a higher mass
inflow rate ($\dot{\cal{M}}_{\rm in}$).  The higher $|v_r|$ corresponds
to a stronger radial neutral--ion drag (since the ions and field remain
pinned at the midplane under the assumption that $\epsilon_{\rm B}=0$;
see footnote~\ref{ansatz}), and therefore (by the ion radial
force-balance condition; see WK93) to a stronger bending of the magnetic
field lines away from the vertical (reflected in the higher value of
$B_{r,{\rm b}}$). This, in turn, increases the level of magnetic
squeezing (reflected in the lower value of $h$), resulting in a stronger
density stratification. Although the thickness of the disc remains
virtually unchanged (at $z_{\rm b} = 1.25\, h_{\rm T}$) and the height
of the sonic point actually goes down in the combined solution (from
$2.90\, h_{\rm T}$ to $2.82\, h_{\rm T}$), the increased stratification
(measured by $z_{\rm b}/h$ and $z_{\rm s}/h$, respectively) means that
the density at the top of the disc is a lower fraction of the midplane
density than in the pure wind solution, leading to a lower wind outflow
rate (measured by $\kappa$) and a correspondingly lower vertical torque
(measured by $T_z$).

The results shown in Fig. \ref{fig:mri} illustrate the possibility that 
the incorporation of MRI-induced
turbulence could increase the overall angular momentum
transport in the disc, but {\em reduce} the amount of
angular momentum carried away by the wind. It may be worth
noting in this connection that the enhanced squeezing of the
disc brought about by the addition of radial transport also
reduces the width of the turbulent layer in comparison with the
value of $z_{\rm t}$ that would have been estimated from the
pure wind solution (in which $a$ increases less rapidly with
$z$). These considerations suggest that the joint operation of
the two angular-momentum transport mechanisms might be self-limiting,
in which case it would be unlikely to introduce a dynamical
instability. This conclusion could be tested through an explicit
stability analysis (e.g. K\"onigl 2004 and references therein)
or by means of numerical simulations.

\begin{table}
\caption{Comparison of the two solutions shown in Fig.~\ref{fig:mri}.}
\begin{tabular}{|c|c|c|}
\hline
Disc/outflow & Pure wind & Wind -- MRI\\
properties  & solution & solution \\
\hline
$a_o$ & $0.55$ & $0.55$ \\
$\epsilon$ & $2.30$ & $2.76$ \\
$h/h_{\rm T}$ & $0.31$ & $0.29$ \\
$z_{\rm t}/h$ & --- & $1.01$ \\
$z_{\rm b}/h$ & $3.99$ & $4.35$ \\
$z_{\rm s}/h$ & $9.25$ & $9.81$ \\    
$\rho_s/\rho_0$ & $1.0\ee-2 $ & $8.3\ee -3 $ \\
$\kappa$ & $0.17$ & $0.14 $ \\
$\lambda$ & $9.60$ & $11.38$ \\
$\xi_{\rm b}^{\prime} \equiv B_{r,{\rm b}}/B_z$ & $1.75$ & $1.81$ \\
$|B_{\phi,b}|/B_z$ & $1.44$ & $1.43$ \\
$T_z$ & $1.63$ & $1.59$ \\
$T_r$ & --- & $0.09$ \\
$\dot{\cal{M}}_{\rm in}$ & $1.74$ & $1.84$ \\   
\hline                                
\end{tabular}
	\label{table:boundary}
\end{table}

\section{Conclusion}
\label{sec:conclusion}

We have constructed a semi-analytic scheme for modeling
magnetized protostellar accretion discs. The model takes into
account both vertical angular-momentum transport associated with
a large-scale magnetic field that threads the disc and drives a
centrifugal wind from its surfaces and radial angular-momentum
transport through MRI-induced turbulence. A novel feature of
this model is an explicit prescription that identifies the
vertical extent of the disc region that is susceptible to
sustainable MRI turbulence at any given radius and incorporates
a turbulent-stress term (based on published results of numerical
simulations and related to the local magnitude of the
large-scale field) into the angular-momentum conservation
relation for this region. Our general scheme employs a
tensor-conductivity formalism and incorporates a realistic
vertical ionization structure for the disc. However, in this
paper we specialize to the ambipolar diffusion-dominated regime
and assume that the ion density is nearly constant with
height. This allows us to compare our results with the pure
vertical-transport solutions obtained for this case by WK93 and
to use the algebraic relations derived in that paper in the
analysis of the combined vertical and radial transport.

We have concentrated on the case where the neutral--ion coupling
parameter $\eta$ is $\gtrsim 1$, which allows both the wind-driving and
MRI-turbulence mechanisms to operate efficiently. Vertical transport
through a wind is expected to dominate when the field-strength parameter
$a$ (the ratio of the local Alfv\'en and sound speeds) is sufficiently
high ($\lesssim 1$), whereas radial MRI-induced transport should prevail
for $a\ll 1$. We constructed a low-$a$ disc solution exhibiting
oscillatory field and velocity profiles and suggested that it evidently
represents the steady-state analogue in a diffusive (and stratified)
disc of the nonlinear MRI channel mode originally identified by Goodman
\& Xu (1994). Although the question of the stability of stratified
channel solutions in the non-ideal MHD regime has not yet been
explicitly addressed, the current evidence (from linear analyses and
numerical simulations) suggests that they may well be unstable. On the
assumption that, in fact, such flows rapidly become turbulent, we
incorporated a turbulent-stress prescription (based on the results of
numerical simulations) into the angular momentum conservation equation
for this region. We showed that, when implemented, this prescription
results in an accretion-disc solution featuring pure inflow. We
argued that radial angular-momentum transport terminates when $2 \eta
a^2$ increases above $\sim 1$, and we presented a hybrid solution in
which this transition occurs at a height $z_{\rm t}$ below the disc
surface ($z_{\rm b}$), with radial transport (restricted to the layer
$z\lesssim z_{\rm t}$) augmenting the vertical transport (which operates
for all $z\le z_{\rm b}$). We found that the joint operation of these
two mechanisms at the same radial location in the disc is apparently
self-limiting. In particular, we showed that the incorporation of radial
transport, while increasing the total amount of angular momentum removed
from the inflowing gas, leads to a reduction in the angular momentum
flux carried away by the wind.

Our results strongly indicate that the two angular-momentum transport
mechanisms are unlikely to have a significant spatial overlap under
realistic circumstances. This is because of the stringent constraints
that must be satisfied for this to occur: the parameter $a$ must be
small enough ($<(2\eta)^{-1/2}$) near the midplane to allow MRI
turbulence to develop over a significant vertical extent, but it cannot
be so small that the field in too weak to launch a wind (or else that
the surface layers become sub-Keplerian). In practice, these constraints
require moderate midplane values of $a$ ($a_0 \gtrsim 0.5$) and rather
low (and slowly decreasing with height) values of $\eta$
($\eta_0\lesssim 1$), which are unlikely to be attained over measurable
radial extents in real discs. If this result indeed applies generally to
protostellar discs in the ambipolar diffusion regime then it is of
interest to inquire where each of these mechanisms is likely to dominate
in such systems.\footnote{It is, however, worth pointing out again in this
connection that vertical angular-momentum transport mediated by
large-scale fields could take place (e.g. in the form of magnetic
braking) even if the conditions for wind launching are not fulfilled.}
As an example, we use the normalizations of $a_0$ and $\eta_0$ given (on
a scale of $\sim 100$\;AU) by equations (3.50) and (3.51), respectively,
of WK93 and the radial scalings ($B\propto r^{-5/4}$, $\rho \propto
r^{-3}$, ion density $\rho_{\rm ion}\propto \rho^{1/2}$) corresponding
to the asymptotic rotational core-collapse solution of Krasnopolsky \&
K\"onigl (2002), which imply that $2 \eta_0 a_0^2 \propto r$ drops below
1 at $r \sim 40$\;AU (with disc winds dominating the angular momentum
transport at larger radii and MRI-induced turbulence turning on closer
in).

The above example should, of course, be taken only as an illustration,
since the radial parameter profiles in actual discs may be quite
different. Furthermore, as the density and column density in the disc
increase with decreasing radius, the Hall conductivity regime, and
eventually also the Ohmic regime, will be encountered (e.g. Balbus \&
Terquem 2001), with different regimes expected to dominate at different
heights at any given radial location (Wardle 2004). We are currently in
the process of applying our general scheme to the investigation of the
various conductivity regimes in realistic protostellar disc models. The
results will be reported in future publications.

\section*{Acknowledgments}

We thank the referee for helpful comments.
This research was supported in part by a NASA Theoretical Astrophysics Program 
grant NNG04G178G and by the Australian Research Council.

\bsp
\label{lastpage}
\end{document}